# An expression for estimating the number of atoms displaced during the irradiation of monolayer graphene with neutrons.

Daniel Codorniu-Pujals, Armando Bermúdez-Martínez

Instituto Superior de Tecnologías y Ciencias Aplicadas (InSTEC). La Habana. Cuba.

e-mail: dcodorniu@instec.cu

The existing mathematical expressions, used to evaluate the number of atoms displaced by particle irradiation of materials, are not applicable to 2D structures like graphene. In a previous work (J. Radioanal. Nucl.Chem, 2011, 289,1, 167-172) a model was developed for performing such evaluations in the case of graphene irradiated with heavy ions and with light charged particles. In the present paper we report the generalization of the model to the graphene irradiated with neutrons in the energy range 0- 2 MeV.

## Introduction.

The irradiation of monolayer graphene could be an effective method for modifying its electronic structure and for achieving specific physical properties adjusted to different applications. For this reason, several groups around the world are working on the irradiation of graphene and a number of results have been already reported [1, 2, 3].

A difficulty arising during planning and studying the irradiation in this system is that most of the models of interaction of the radiation with matter cannot be applied to the two-dimensional structure of graphene. In particular, the mathematical expressions available to calculate the number of atoms displaced during the bombardment with particles can be applied only to 3D isotropic solids. In a previous work of our group [4] an alternative analytic expression was presented for the irradiation of graphene with heavy ions, protons and other light charged particles. The expression was obtained on the basis of the classic theory of scattering, using a Coulomb potential for the light charged particles and an Inverse Square potential for heavy ions.

In the present paper we have extended the model to include neutrons with energies below 2 MeV. In this energy range there is a simple linear dependence of the cross section of neutron scattering by carbon atoms with the energy of neutrons [5] and , then, it is valid to apply the classic theory of scattering. The region above 2 MeV is out of the scope of this work. In this region resonance effects are present and the analysis requires a quantum-mechanical approach. From the practical point of view, the range from 0 to 2 MeV has particular interest because it covers the energy values of the neutrons produced in most of the nuclear reactors.

## Application of the model to the case of neutrons.

Following the procedure developed in [4] we have started with the general expression for the number of atoms of the lattice displaced by each incident neutron with energy $E$:

$$c_D = W(E)[1 + k(E)P_{pl}(E)] \quad (1)$$

where $W(E)$ is the number of atoms displaced directly by the incident particle, i.e. the number of Primary Knock-on Atoms (PKA); $P_{pl}(E)$ is the probability that a PKA comes out in a direction very near to the crystalline plane, so it is able to displace other atoms[1]; and $k(E)$ is the average number of secondary atoms displaced by each PKA.

The expression (1) is valid if $E > E_{umb}$, where $E_{umb}$ is the minimum energy of the incident neutron able to create a PKA. This energy corresponds to the minimum kinetic energy, $T_d$, that a carbon atom should acquire in order to escape from its equilibrium position. There is a well-known relationship between $E_{umb}$ and $T_d$:

$$E_{umb} = T_d(M_1 + M_2)^2 / 4M_1M_2,$$

where $M_1$ and $M_2$ are the masses of the incident particles and of the carbon atoms, respectively [6]. According with the literature, the value of $T_d$ for graphene is about 15eV [1], then we can estimate $E_{umb}$ for neutrons in graphene as 50-60eV.

For $E < E_{umb}$, $c_D = 0$.

$P_{pl}(E)$ is calculated from the expression:

$$P_{pl}(E) = \left(\frac{2\pi}{\sigma_T}\right) \int_{\phi_m}^{\phi_c} \sigma(\phi)\sin\phi \, d\phi \quad (2)$$

Where $\phi$ is the angle, formed by the exit direction of the PKA with the direction of the ingoing particle in the system of the center of masses (CM). We define $\theta$ as the angle between the incident beam and the normal to the plane of the crystal and $\beta$ as the angle between the direction of exit of the PKA and the plane of the crystal in the laboratory system. These three angles are related by the relationship $\phi = 2(\theta + \beta)$ [7]. $\sigma(\phi)$ is the cross section for the exit of the PKA in an angle between $\phi$ and $\phi + d\phi$, and $\sigma_T$ the total cross section for the exit in any angle in the system of the CM. $\phi_c$ and $\phi_m$ are the scattering angles corresponding to $\beta = \beta_c$ and $\beta = 0$, respectively.

Considering that the neutrons are electrically neutral we can neglect their interactions with the electrons of the carbon atoms and take into account only the interaction with their nuclei. Usually this interaction is represented by a hard spheres potential [6]:

$$V(r) = \begin{cases} \infty & r > a \\ 0 & r < a \end{cases} \quad (3)$$

where $a$ is the radius of the carbon nucleus.

---

[1] It is just the probability that the PKA goes out with an angle (relative to the plane of the crystal) lesser than a critical value, $\beta_c$, above which the PKA escapes from the system. According to the estimation made in [4] this angle is around 0.2 radians for graphene.

For this kind of potential the cross sections are given by [3]:

$$\sigma(\phi) = \frac{a^2}{4} \text{ and } \sigma_T = \pi a^2$$

Using these expressions in formula (2) and integrating, we obtain:

$$P_{pl}(E) = \frac{1}{2}[cos2\theta + cos(2\theta + 2\beta_c)] \quad (4)$$

The expression for the average number of secondary atoms displaced by each PKA, $k(E)$, is obtained in exactly the same way that it was done for heavy ions and light charged particles in [4], then we have:

$$k(E) = \begin{cases} 0 & E < E_{umb}^{pl} \\ \frac{1}{2}\sqrt{\frac{E}{E_{umb}}}\sin(\beta_c + \theta) & E > E_{umb}^{pl} \end{cases} \quad (5)$$

,where $E_{umb}^{pl} = \frac{E_{umb}}{sin^2(\beta_c+\theta)}$ is the energy threshold corresponding to those atoms able to create collision cascades in the graphene plane.

With the objective of determining the number of primary atoms displaced by each incident particle, $W(E)$ ,we should have in mind that, in the case of neutrons, the probability of displacement of two or more carbon atoms in a single collision process can be taken as zero .

On the other hand, it is necessary to consider the possibility of channeling of the incident particles through the lattice interstices.

Then, we can write:

$$W(E) = \begin{cases} 0 & E < E_{umb} \\ \mu & E > E_{umb} \end{cases} \quad (6)$$

,where $\mu$ is a parameter that should be determined experimentally. It is obvious that $0<\mu<1$. Furthermore, we can determine more precisely the order of magnitude of $\mu$ by the following considerations:

Since the interaction of the neutrons is mainly with the nuclei of the carbon atoms and not with their electronic clouds, and following the pattern of "hard spheres", the probability that a neutron meets a carbon nucleus will be of the order of the ratio between the space occupied by the nuclei ($\sim 10^{-15}m$ ) and the interatomic space ($\sim 10^{-10}m$ ), so that , the order of magnitude of $\mu$ will be $\sim 10^{-5}m$ . This is an important difference with the case of ions, studied in [1] , where $\mu \sim 1$.

Finally, combining equations (1), (4), (5) and (6) we get the expression for the number of atoms of the lattice displaced by each incident neutron (with energy $E < 2MeV$):

$$c_D = \begin{cases} 0 & E < E_{umb}^{pl} \\ \mu & E_{umb}^{pl} < E < E_{umb} \\ \mu\left[1 + \frac{1}{4}\sqrt{\frac{E}{E_{umb}}}\sin(\beta_c + \theta)[\cos 2\theta + \cos(2\theta + 2\beta_c)]\right] & E_{umb} < E < 2MeV \end{cases} \qquad (7)$$